\documentclass{nature}
\usepackage{graphicx} 
\usepackage{cite}
\usepackage{epsfig} 
\usepackage{xcolor}

\title{Topologically Enabled Ultra-high-Q Guided Resonances Robust to Out-of-plane Scattering}

\author{Jicheng Jin$^{1}$, Xuefan Yin$^1$, Liangfu Ni$^1$, Marin Solja\v{c}i\'{c}$^2$, Bo Zhen$^3$, \& Chao Peng$^{1,*}$}
\begin{document}
\maketitle

\begin{affiliations}
 \item State Key Laboratory of Advanced Optical Communication Systems and Networks, Peking University, Beijing 100871, China
   \item Department of Physics, Massachusetts Institute of Technology, Cambridge, MA 02139, USA.
 \item Department of Physics and Astronomy, University of Pennsylvania, Philadelphia, PA 19104, USA

\end{affiliations}

\begin{abstract}

Due to their ability to confine light, optical resonators\cite{biberman2012ultralow,hossein2007free,akahane_high-q_2003} are of great importance to science and technology, yet their performances are often limited by out-of-plane scattering losses from inevitable fabrication imperfections\cite{minkov_statistics_2013,ishizaki_numerical_2009}. Here, we theoretically propose and experimentally demonstrate a class of guided resonances in photonic crystal slabs, where out-of-plane scattering losses are strongly suppressed due to their topological nature. Specifically, these resonances arise when multiple bound states in the continuum - each carrying a topological charge\cite{zhen_topological_2014} - merge in the momentum space and enhance the quality factors of all resonances nearby. We experimentally achieve quality factors as high as $4.9\times 10^5$ based on these resonances in the telecommunication regime, which is 12-times higher than ordinary designs.
We further show this enhancement is robust across the samples we fabricated.
Our work paves the way for future explorations of topological photonics in systems with open boundary condition and their applications in improving optoelectronic devices in photonic integrated circuits.
\end{abstract}

Topological defects\cite{mermin_topological_1979} are ubiquitous in nature. Examples range from quantum vortices in superfluids to singular optical beams\cite{gbur_singular_2016}, which are characterized by the non-trivial winding patterns of system parameters (velocity, phase, or polarization) in real space.
Recently it is found that unexpected topological defects can also emerge in the momentum space of a crystal and give rise to interesting physical consequences: one such example is the optical bound states in the continuum (BICs). BICs reside inside the continuous spectrum of extended states, yet, defying the common intuition, remain perfectly localized in space and their lifetimes are supposed to be infinitely long.
Since their initial proposal\cite{von_neuman_uber_1929}, BICs have been observed in a variety of wave systems, including photonic\cite{imada1999coherent,hsu_bound_2016,hsu_observation_2013,plotnik_experimental_2011,fan_analysis_2002,kodigala_lasing_2017,gansch_measurement_2016,gomis-bresco_anisotropy-induced_2017,molina_surface_2012,carletti_giant_2018,friedrich_interfering_1985,monticone2014embedded}, phononic\cite{lim_character_1969}, and water waves\cite{cobelli_experimental_2009}.
In photonic crystal (PhC) slabs, their fundamental nature has been identified to be topological: they are essentially topological defects of polarization directions defined in the momentum space\cite{zhen_topological_2014}. In practice\cite{hsu_observation_2013,lee2014fabricating}, the $Q$s of BICs often fall much shorter of their theoretical prediction of infinity, limited to only about $1 \times 10^4$. Aside from other contributing factors such as material absorption or the finite size of the samples, the main limiting factor of the $Q$ of BICs comes from out-of-plane scattering losses from fabrication imperfections or disorder - a common problem shared among many high-$Q$ on-chip resonators~\cite{hughes_extrinsic_2005,minkov_statistics_2013,ishizaki_numerical_2009,biberman2012ultralow,hossein2007free}.

Here we theoretically propose and experimentally demonstrate on-chip photonic resonances that are much less susceptible to out-of-plane scattering losses than usual due to their unique topological nature.
Specifically, we first show that the topological charges of BICs control the $Q$s of their surrounding resonances; more importantly, when multiple BICs are designed to merge, all modes nearby enjoy significant enhancements of their $Q$s due to a modified scaling rule.
We further numerically show that the resulting resonances, in this new topological configuration, become robust to fabrication imperfections and disorder. Finally, we experimentally demonstrate a 12-times enhancement of $Q$s in fabricated samples using this new topological configuration of BICs over previous designs.

We start by showing that resonances with ultra-high quality factors ($Q$s) that are much more robust
to out-of-plane scattering from disorder can be achieved by merging multiple topological charges
carried by BICs.
First, we consider a PhC slab (Fig. 1a), where a square lattice (periodicity $a = 519.25$ nm) of circular air holes (radius $r = 175 $ nm) is patterned in a silicon layer (thickness of $h=600$ nm) placed in the air.
Through numerical simulations (COMSOL Multiphysics), we focus on the lowest TE-like band in the continuum (TE-A, red line) whose lifetime goes to infinity at 9 discrete $k$ points as 9 BICs (top left panel of Fig. 1b).
The topological nature of the BICs can be understood from the corresponding far-field polarization plot (bottom left panel), where each BIC appears as a topological defect (vortex) of polarization long axes\cite{zhen_topological_2014,zhou_observation_2018,bulgakov_topological_2017,bulgakov_bound_2017,iwahashi_higher-order_2011,kitamura_focusing_2012,zhang_observation_2018} characterized by an integer topological charge of $\pm 1$.
Among these 9 vortices, one is pinned at the center of the Brilluion zone (BZ) due to symmetry, while the locations of rest 8 can be controlled by varying system parameters such as periodicity $a$.
For example, when $a$ increases from $519.25$ nm to $531.42$ nm, the 8 off-center vortices move towards the center before all 9 of them merge into a single BIC with charge of $+1$ as $a$ further increases to $580$ nm.

The topological configuration of BICs controls the radiation loss of all nearby resonances, which further determines the highest $Q$ achievable in practice as shown later.
Specifically, $Q$ is shown to scale quadratically ($\propto 1/k^2)$ as the distance ($k$) away from a single isolated BIC with charge $\pm 1$; however, this scaling changes to $Q \propto 1/k^6$ in a sample where all 9 BICs just merge (noted as the ``merging-BIC design" hereafter). The comparison between these two scenarios are shown in Fig. 1c, where $Q$s in a merging-BIC design (red) are always orders-of-magnitude higher than those in an isolated-BIC design (blue) along all directions in $k$ space due to their fundamentally different scaling properties. This difference in scaling originates from the different asymptotic behaviors of radiation amplitudes $\sqrt{1/Q}$: in the isolated-BIC case, $\sqrt{1/Q} \propto k$; in comparison, when there are also off-center BICs at $\pm k_{\mathrm{BIC}}$, $\sqrt{1/Q}$ becomes proportional to $(k + k_{\mathrm{BIC}})(k - k_{\mathrm{BIC}})k$. In the merging-BIC design, $k_{\mathrm{BIC}} = 0$ and we get $1/Q \propto k^6$. This different scaling is similar to some of the recent findings\cite{yuan2018bound,bulgakov2017bound,yuan_strong_2017}.
Further explanations from the viewpoint of coupled-wave theory is presented in section I and II of the Supplementary Information.

While simulation results of infinitely-large perfect PhCs set the theoretical upper bounds of $Q$s, realistic samples (schematically shown in Fig. 2a) feature a few major differences that govern the highest $Q$ achievable in practice.
First, all samples are finite in size; their boundaries break the translation symmetry and introduce fractional orders of the primitive reciprocal lattice in $k$ space (green dots in Fig. 2a)\cite{liang_three-dimensional_2012,wang_mode_2016}.
As a result, each infinitely-large Bloch mode with a single $k$-component is split into a series of finite-size modes with multiple $k$-components.
See Supplementary Information Section \uppercase\expandafter{\romannumeral3} for an example of this effect experimentally observed in our sample.
Second, all fabricated samples exhibit disorder and imperfections with both long- and short-range correlations, allowing modes at different $k$ points to couple to each other. Due to these inevitable coupling terms, modes at different fractional momentum orders are hybridized and all of their loss channels become available to the final resonance\cite{ni_analytical_2017}.

The advantage of our merging-BIC design over an isolated-BIC design is confirmed in our simulation results (COMSOL Multiphysics) of perturbed $15 \times 15$ PhC super-cells.
In a perfect super-cell structure without disorder, the BIC mode with infinite-$Q$ remains at the center of the BZ (Fig. 2b, upper panel).
To compare, perturbations are applied to both the radii ($\Delta r$) and positions of the holes ($\Delta x, \Delta y$) following the statistics that best captures our samples described later in Fig. 3.
As expected, each mode in disordered samples exhibits multiple components in the $k$ space.
Furthermore, resonances in a disordered sample originated in a
merging-BIC design have significantly lower radiation fields than those from a isolated-BIC design with the same disorder (Fig. 2c).
This result agrees well with Fig. 1b,c: all modes contributing to the final resonance in the merging-BIC sample have much higher $Q$s than those in the isolated-BIC case; naturally, resonances in the former sample are much more immune to out-of-plane scattering from disorder than the latter.
Finally, this enhancement of $Q$ is observed to be robust across a range of $k$ as shown in the quantitative comparison (Fig. 2d). {Here, all holes are asymmetric to present typical fabrication error with tilted angle $\theta \approx$2$^\circ$ and center shift  ($\Delta x$=2nm, $\Delta y$=4nm) before applying disorder. (see Supplementary Information section \uppercase\expandafter{\romannumeral4} and \uppercase\expandafter{\romannumeral5} for details)}

To  verify our theoretical findings, we fabricate PhC samples with both merging-BIC and isolated-BIC designs using the same e-beam lithography (EBL) and induced coupled plasma (ICP) etching processes on a 600 nm thick silicon-on-insulator wafer (see Methods for details). The underlying SiO$_2$ layer is then removed to restore the up-down mirror symmetry required by tunable BICs \cite{hsu_observation_2013,zhen_topological_2014}. The samples are about 250  $\times$ 250 $\mu$m in size. The periodicity of the sample is varied from 530 to 580 nm to sample through designs with merging and isolated BICs. From the scanning electron microscope images of the samples (Fig. 3a,b), the standard deviations of hole locations and radii are estimated to be about 5 nm, which is applied to the numerical simulations presented above.

 The experimental setup is schematically shown in Fig.3c. A tunable telecommunication laser in the C-band is first sent through a X-polarizer (Pol,X) before being focused by lens 1 (L1) onto the back focal plane of an infinity-corrected objective lens. The incident angle of the laser on the sample is thus controlled by moving L1 in the $x-y$ plane. Through this confocal setup, reflected and scattered light are also collected by the same objective; they are then expanded by 1.67 times through a relay 4-f system and imaged on a camera. A Y-polarizer (Pol, Y) is used to block reflected light (X-polarized), while allowing scattered light to pass (see Supplementary Information Section \uppercase\expandafter{\romannumeral6} for details). Under the on-resonance coupling condition, where the PhC sample supports a resonance at the same wavelength as the incident light at the incident angle, iso-frequency contours are observed on the camera, similar to previously reported results \cite{regan_direct_2016,zhou_observation_2018}. Three examples of isofrequency contours are schematically shown in Fig. 4a as dashed lines.

The quality factors of resonances at different $k$ points are further characterized through scattered light. Specifically, a movable pin hole (not shown in Fig. 3c) is placed at the image plane of the rear focal plane of the objective to specify a $k$ point. A photo-diode connected to a lock-in amplifier is placed behind the pin hole to record scattered light intensity as a function of the tunable laser wavelength (see Supplementary Information Section \uppercase\expandafter{\romannumeral6} for details). As shown in Fig. 4a, when different $k$ points are selected by the pin hole (X, Y, and Z), different scattering spectra are observed while all exhibiting symmetric Lorentzian features. Similar scattering phenomena have been observed before\cite{regan_direct_2016}, and can be understood as the follows: scattered light intensity is governed by the spectral density of states of the sample at this $k$ point, which are Lorentzian functions centered at the resonance frequencies with linewidths determined by the $Q$ of the resonances (see Supplementary Information Section \uppercase\expandafter{\romannumeral7} for details).

The quality factors of the resonances are extracted by numerically fitting the scattering spectra to Lorentzian functions. As shown in Fig. 4a, $Q$ increases from $2.6 \times 10^5$ to $4.5 \times 10^5$ as the observing point moves closer to the center of the BZ from X to Z. This agrees well with simulation results in Fig. 1. The highest $Q$ observed on the merging-BIC sample is $4.9 \times 10^5$ at point W (Fig. 4b). In comparison, the highest $Q$ observed on the isolated-BIC sample, fabricated on the same wafer through the same processes only with different structural parameters, is limited to only $4 \times 10^4$ - over an order of magnitude lower (Fig. 4c). This confirms our simulation results in Fig. 2 that engineering the topological configurations of BICs can significantly suppress scattering losses. Furthermore, this over-ten-fold enhancement of quality factor is observed to be robust: not only does it appear over a wide range in the $k$ space as shown in Fig. 5, similar level of enhancements also appears in all merging-BIC samples we fabricated (see Supplementary Information Section \uppercase\expandafter{\romannumeral7} for details).

Topological photonics\cite{lu_topological_2014,ozawa2018topological,wang_observation_2009} have found tremendous success in suppressing in-plane back-scattering losses, often based on topological protections in non-reciprocal systems with broken time-reversal symmetry. Here, we focus on a different class of problems: to suppress the out-of-plane scattering losses in a reciprocal system using concepts from topology. By merging multiple topological charges carried by BICs, we experimentally demonstrate PhC resonances with record-high quality factors of $Q = 4.9 \times 10^5$, over an order of magnitude higher than ordinary designs. These ultra-high-Q resonances are potentially useful for chemical or biological sensing \cite{luchansky2011high,di2009chemical}, nonlinear generation\cite{logan2018400}, and large-area laser applications\cite{hirose2014watt}.
Furthermore, governed by their topological nature, these high-$Q$ resonances are observed to be robust against fabrication imperfections, which paves the way to improve the performance of optoelectronic devices using concepts from topological photonics.
Finally, our fundamental concept of topological-defect engineering holds for general linear wave systems, ranging from photonics to acoustics and electronics.

\begin{methods}
\subsection{Sample fabrication.}
The sample was fabricated on a silicon-on-insulator (SOI) wafer with e-beam lithography (EBL) followed by induced coupled plasma (ICP) etching. For EBL, the SOI wafer was spin-coated with a 330nm-thick layer of ZEP520A photo-resist before being exposed with EBL (JBX-6300FS) at beam current of $400$ pA and field size of 500 $\mu$m. The sample was then etched with ICP (Oxford Plasmapro Estrelas 100) using a mixture of SF$_6$ and C$_4$F$_8$. After etching, the resist was removed with N-Methyl-2-pyrrolidone and the buried oxide layer was removed using $49\%$ HF.

\subsection{Measurement system.} The incident light source was a tunable C-band telecommunication laser (Santec TSL-550), which was sent through a chopper for lock-in detection. A pin hole with diameter of 500 $\mu$m was placed on the Fourier plane to pick out desired wavevectors. Scattered light through the pin hole was collected by a photo-diode (PDA10DT-EC), which was connected to a lock-in amplifier (SRS SR830). A flip mirror was used to switch between the camera that image iso-frequency contours and the photo-diode. Besides characterizing far-field radiation patterns, the setup could also take near-field images of the sample if another lens was inserted into the optical path.

\end{methods}


\bibliography{reference}{}

\begin{thebibliography}{10}
\expandafter\ifx\csname url\endcsname\relax
  \def\url#1{\texttt{#1}}\fi
\expandafter\ifx\csname urlprefix\endcsname\relax\def\urlprefix{URL }\fi
\providecommand{\bibinfo}[2]{#2}
\providecommand{\eprint}[2][]{\url{#2}}

\bibitem{biberman2012ultralow}
\bibinfo{author}{Biberman, A.}, \bibinfo{author}{Shaw, M.~J.},
  \bibinfo{author}{Timurdogan, E.}, \bibinfo{author}{Wright, J.~B.} \&
  \bibinfo{author}{Watts, M.~R.}
\newblock \bibinfo{title}{Ultralow-loss silicon ring resonators}.
\newblock \emph{\bibinfo{journal}{Optics letters}}
  \textbf{\bibinfo{volume}{37}}, \bibinfo{pages}{4236--4238}
\newblock  (\bibinfo{year}{2012}).

\bibitem{hossein2007free}
\bibinfo{author}{Hossein-Zadeh, M.} \& \bibinfo{author}{Vahala, K.~J.}
\newblock \bibinfo{title}{Free ultra-high-q microtoroid: a tool for designing
  photonic devices}.
\newblock \emph{\bibinfo{journal}{Optics Express}}
  \textbf{\bibinfo{volume}{15}}, \bibinfo{pages}{166--175}
\newblock  (\bibinfo{year}{2007}).

\bibitem{akahane_high-q_2003}
\bibinfo{author}{Akahane, Y.}, \bibinfo{author}{Asano, T.},
  \bibinfo{author}{Song, B.-S.} \& \bibinfo{author}{Noda, S.}
\newblock \bibinfo{title}{High-\textit{{Q}} photonic nanocavity in a
  two-dimensional photonic crystal}.
\newblock \emph{\bibinfo{journal}{Nature}} \textbf{\bibinfo{volume}{425}},
  \bibinfo{pages}{944--947}
\newblock  (\bibinfo{year}{2003}).

\bibitem{minkov_statistics_2013}
\bibinfo{author}{Minkov, M.}, \bibinfo{author}{Dharanipathy, U.~P.},
  \bibinfo{author}{Houdré, R.} \& \bibinfo{author}{Savona, V.}
\newblock \bibinfo{title}{Statistics of the disorder-induced losses of high-{Q}
  photonic crystal cavities}.
\newblock \emph{\bibinfo{journal}{Optics Express}}
  \textbf{\bibinfo{volume}{21}}, \bibinfo{pages}{28233--28245}
\newblock  (\bibinfo{year}{2013}).

\bibitem{ishizaki_numerical_2009}
\bibinfo{author}{Ishizaki, K.}, \bibinfo{author}{Okano, M.} \&
  \bibinfo{author}{Noda, S.}
\newblock \bibinfo{title}{Numerical investigation of emission in finite-sized,
  three-dimensional photonic crystals with structural fluctuations}.
\newblock \emph{\bibinfo{journal}{Journal of the Optical Society of America B}}
  \textbf{\bibinfo{volume}{26}}, \bibinfo{pages}{1157--1161}
\newblock  (\bibinfo{year}{2009}).

\bibitem{zhen_topological_2014}
\bibinfo{author}{Zhen, B.}, \bibinfo{author}{Hsu, C.~W.}, \bibinfo{author}{Lu,
  L.}, \bibinfo{author}{Stone, A.~D.} \& \bibinfo{author}{Solja\v{c}i\'{c}, M.}
\newblock \bibinfo{title}{Topological {nature} of {optical} {bound} {states} in
  the {continuum}}.
\newblock \emph{\bibinfo{journal}{Physical Review Letters}}
  \textbf{\bibinfo{volume}{113}}, \bibinfo{pages}{257401}
\newblock  (\bibinfo{year}{2014}).

\bibitem{mermin_topological_1979}
\bibinfo{author}{Mermin, N.~D.}
\newblock \bibinfo{title}{The topological theory of defects in ordered media}.
\newblock \emph{\bibinfo{journal}{Reviews of Modern Physics}}
  \textbf{\bibinfo{volume}{51}}, \bibinfo{pages}{591--648}
\newblock  (\bibinfo{year}{1979}).

\bibitem{gbur_singular_2016}
\bibinfo{author}{Gbur, G.~J.}
\newblock \emph{\bibinfo{title}{Singular {Optics}}}
\newblock  (\bibinfo{publisher}{CRC Press}, \bibinfo{year}{2016}).

\bibitem{von_neuman_uber_1929}
\bibinfo{author}{von Neuman, J.} \& \bibinfo{author}{Wigner, E.}
\newblock \bibinfo{title}{Uber merkwürdige diskrete {Eigenwerte}. {Uber} das
  {Verhalten} von {Eigenwerten} bei adiabatischen {Prozessen}}.
\newblock \emph{\bibinfo{journal}{Physikalische Zeitschrift}}
  \textbf{\bibinfo{volume}{30}}, \bibinfo{pages}{467--470}
\newblock  (\bibinfo{year}{1929}).

\bibitem{imada1999coherent}
\bibinfo{author}{Imada, M.} \emph{et~al.}
\newblock \bibinfo{title}{Coherent two-dimensional lasing action in
  surface-emitting laser with triangular-lattice photonic crystal structure}.
\newblock \emph{\bibinfo{journal}{Applied physics letters}}
  \textbf{\bibinfo{volume}{75}}, \bibinfo{pages}{316--318}
\newblock  (\bibinfo{year}{1999}).

\bibitem{hsu_bound_2016}
\bibinfo{author}{Hsu, C.~W.}, \bibinfo{author}{Zhen, B.},
  \bibinfo{author}{Stone, A.~D.}, \bibinfo{author}{Joannopoulos, J.~D.} \&
  \bibinfo{author}{Solja\v{c}i\'{c}, M.}
\newblock \bibinfo{title}{Bound states in the continuum}.
\newblock \emph{\bibinfo{journal}{Nature Reviews Materials}}
  \textbf{\bibinfo{volume}{1}}, \bibinfo{pages}{16048}
\newblock  (\bibinfo{year}{2016}).

\bibitem{hsu_observation_2013}
\bibinfo{author}{Hsu, C.~W.} \emph{et~al.}
\newblock \bibinfo{title}{Observation of trapped light within the radiation
  continuum}.
\newblock \emph{\bibinfo{journal}{Nature}} \textbf{\bibinfo{volume}{499}},
  \bibinfo{pages}{188--191}
\newblock  (\bibinfo{year}{2013}).

\bibitem{plotnik_experimental_2011}
\bibinfo{author}{Plotnik, Y.} \emph{et~al.}
\newblock \bibinfo{title}{Experimental {observation} of {optical} {bound}
  {states} in the {continuum}}.
\newblock \emph{\bibinfo{journal}{Physical Review Letters}}
  \textbf{\bibinfo{volume}{107}}, \bibinfo{pages}{183901}
\newblock  (\bibinfo{year}{2011}).

\bibitem{fan_analysis_2002}
\bibinfo{author}{Fan, S.} \& \bibinfo{author}{Joannopoulos, J.~D.}
\newblock \bibinfo{title}{Analysis of guided resonances in photonic crystal
  slabs}.
\newblock \emph{\bibinfo{journal}{Physical Review B}}
  \textbf{\bibinfo{volume}{65}}, \bibinfo{pages}{235112}
\newblock  (\bibinfo{year}{2002}).

\bibitem{kodigala_lasing_2017}
\bibinfo{author}{Kodigala, A.} \emph{et~al.}
\newblock \bibinfo{title}{Lasing action from photonic bound states in
  continuum}.
\newblock \emph{\bibinfo{journal}{Nature}} \textbf{\bibinfo{volume}{541}},
  \bibinfo{pages}{196--199}
\newblock  (\bibinfo{year}{2017}).

\bibitem{gansch_measurement_2016}
\bibinfo{author}{Gansch, R.} \emph{et~al.}
\newblock \bibinfo{title}{Measurement of bound states in the continuum by a
  detector embedded in a photonic crystal}.
\newblock \emph{\bibinfo{journal}{Light: Science \& Applications}}
  \textbf{\bibinfo{volume}{5}}, \bibinfo{pages}{e16147}
\newblock  (\bibinfo{year}{2016}).

\bibitem{gomis-bresco_anisotropy-induced_2017}
\bibinfo{author}{Gomis-Bresco, J.}, \bibinfo{author}{Artigas, D.} \&
  \bibinfo{author}{Torner, L.}
\newblock \bibinfo{title}{Anisotropy-induced photonic bound states in the
  continuum}.
\newblock \emph{\bibinfo{journal}{Nature Photonics}}
  \textbf{\bibinfo{volume}{11}}, \bibinfo{pages}{232--236}
\newblock  (\bibinfo{year}{2017}).

\bibitem{molina_surface_2012}
\bibinfo{author}{Molina, M.~I.}, \bibinfo{author}{Miroshnichenko, A.~E.} \&
  \bibinfo{author}{Kivshar, Y.~S.}
\newblock \bibinfo{title}{Surface {bound} {states} in the {continuum}}.
\newblock \emph{\bibinfo{journal}{Physical Review Letters}}
  \textbf{\bibinfo{volume}{108}}, \bibinfo{pages}{070401}
\newblock  (\bibinfo{year}{2012}).

\bibitem{carletti_giant_2018}
\bibinfo{author}{Carletti, L.}, \bibinfo{author}{Koshelev, K.},
  \bibinfo{author}{De~Angelis, C.} \& \bibinfo{author}{Kivshar, Y.}
\newblock \bibinfo{title}{Giant {nonlinear} {response} at the {nanoscale}
  {driven} by {bound} {states} in the {continuum}}.
\newblock \emph{\bibinfo{journal}{Physical Review Letters}}
  \textbf{\bibinfo{volume}{121}}, \bibinfo{pages}{033903}
\newblock  (\bibinfo{year}{2018}).

\bibitem{friedrich_interfering_1985}
\bibinfo{author}{Friedrich, H.} \& \bibinfo{author}{Wintgen, D.}
\newblock \bibinfo{title}{Interfering resonances and bound states in the
  continuum}.
\newblock \emph{\bibinfo{journal}{Physical Review A}}
  \textbf{\bibinfo{volume}{32}}, \bibinfo{pages}{3231--3242}
\newblock  (\bibinfo{year}{1985}).

\bibitem{monticone2014embedded}
\bibinfo{author}{Monticone, F.} \& \bibinfo{author}{Alu, A.}
\newblock \bibinfo{title}{Embedded photonic eigenvalues in 3d nanostructures}.
\newblock \emph{\bibinfo{journal}{Physical Review Letters}}
  \textbf{\bibinfo{volume}{112}}, \bibinfo{pages}{213903}
\newblock  (\bibinfo{year}{2014}).

\bibitem{lim_character_1969}
\bibinfo{author}{Lim, T.~C.} \& \bibinfo{author}{Farnell, G.~W.}
\newblock \bibinfo{title}{Character of {pseudo} {surface} {waves} on
  {anisotropic} {crystals}}.
\newblock \emph{\bibinfo{journal}{The Journal of the Acoustical Society of
  America}} \textbf{\bibinfo{volume}{45}}, \bibinfo{pages}{845--851}
\newblock  (\bibinfo{year}{1969}).

\bibitem{cobelli_experimental_2009}
\bibinfo{author}{Cobelli, P.~J.}, \bibinfo{author}{Pagneux, V.},
  \bibinfo{author}{Maurel, A.} \& \bibinfo{author}{Petitjeans, P.}
\newblock \bibinfo{title}{Experimental observation of trapped modes in a water
  wave channel}.
\newblock \emph{\bibinfo{journal}{EPL (Europhysics Letters)}}
  \textbf{\bibinfo{volume}{88}}, \bibinfo{pages}{20006}
\newblock  (\bibinfo{year}{2009}).

\bibitem{lee2014fabricating}
\bibinfo{author}{Lee, J.}, \bibinfo{author}{Zhen, B.}, \bibinfo{author}{Chua,
  S.-L.}, \bibinfo{author}{Shapira, O.} \& \bibinfo{author}{Solja{\v{c}}i{\'c},
  M.}
\newblock \bibinfo{title}{Fabricating centimeter-scale high quality factor
  two-dimensional periodic photonic crystal slabs}.
\newblock \emph{\bibinfo{journal}{Optics express}}
  \textbf{\bibinfo{volume}{22}}, \bibinfo{pages}{3724--3731}
\newblock  (\bibinfo{year}{2014}).

\bibitem{hughes_extrinsic_2005}
\bibinfo{author}{Hughes, S.}, \bibinfo{author}{Ramunno, L.},
  \bibinfo{author}{Young, J.~F.} \& \bibinfo{author}{Sipe, J.~E.}
\newblock \bibinfo{title}{Extrinsic {optical} {scattering} {loss} in {photonic}
  {crystal} {waveguides}: {role} of {fabrication} {disorder} and {photon}
  {group} {velocity}}.
\newblock \emph{\bibinfo{journal}{Physical Review Letters}}
  \textbf{\bibinfo{volume}{94}}, \bibinfo{pages}{033903}
\newblock  (\bibinfo{year}{2005}).

\bibitem{zhou_observation_2018}
\bibinfo{author}{Zhou, H.} \emph{et~al.}
\newblock \bibinfo{title}{Observation of bulk {Fermi} arc and polarization half
  charge from paired exceptional points}.
\newblock \emph{\bibinfo{journal}{Science}} \bibinfo{pages}{eaap9859}
\newblock  (\bibinfo{year}{2018}).

\bibitem{bulgakov_topological_2017}
\bibinfo{author}{Bulgakov, E.~N.} \& \bibinfo{author}{Maksimov, D.~N.}
\newblock \bibinfo{title}{Topological {bound} {states} in the {continuum} in
  {arrays} of {dielectric} {spheres}}.
\newblock \emph{\bibinfo{journal}{Physical Review Letters}}
  \textbf{\bibinfo{volume}{118}}, \bibinfo{pages}{267401}
\newblock  (\bibinfo{year}{2017}).

\bibitem{bulgakov_bound_2017}
\bibinfo{author}{Bulgakov, E.~N.} \& \bibinfo{author}{Maksimov, D.~N.}
\newblock \bibinfo{title}{Bound states in the continuum and polarization
  singularities in periodic arrays of dielectric rods}.
\newblock \emph{\bibinfo{journal}{Physical Review A}}
  \textbf{\bibinfo{volume}{96}}, \bibinfo{pages}{063833}
\newblock  (\bibinfo{year}{2017}).

\bibitem{iwahashi_higher-order_2011}
\bibinfo{author}{Iwahashi, S.} \emph{et~al.}
\newblock \bibinfo{title}{Higher-order vector beams produced by
  photonic-crystal lasers}.
\newblock \emph{\bibinfo{journal}{Optics Express}}
  \textbf{\bibinfo{volume}{19}}, \bibinfo{pages}{11963--11968}
\newblock  (\bibinfo{year}{2011}).

\bibitem{kitamura_focusing_2012}
\bibinfo{author}{Kitamura, K.}, \bibinfo{author}{Sakai, K.},
  \bibinfo{author}{Takayama, N.}, \bibinfo{author}{Nishimoto, M.} \&
  \bibinfo{author}{Noda, S.}
\newblock \bibinfo{title}{Focusing properties of vector vortex beams emitted by
  photonic-crystal lasers}.
\newblock \emph{\bibinfo{journal}{Optics Letters}}
  \textbf{\bibinfo{volume}{37}}, \bibinfo{pages}{2421--2423}
\newblock  (\bibinfo{year}{2012}).

\bibitem{zhang_observation_2018}
\bibinfo{author}{Zhang, Y.} \emph{et~al.}
\newblock \bibinfo{title}{Observation of {polarization} {vortices} in
  {momentum} {space}}.
\newblock \emph{\bibinfo{journal}{Physical Review Letters}}
  \textbf{\bibinfo{volume}{120}}, \bibinfo{pages}{186103}
\newblock  (\bibinfo{year}{2018}).

\bibitem{yuan2018bound}
\bibinfo{author}{Yuan, L.} \& \bibinfo{author}{Lu, Y.~Y.}
\newblock \bibinfo{title}{Bound states in the continuum on periodic structures
  surrounded by strong resonances}.
\newblock \emph{\bibinfo{journal}{Physical Review A}}
  \textbf{\bibinfo{volume}{97}}, \bibinfo{pages}{043828}
\newblock  (\bibinfo{year}{2018}).

\bibitem{bulgakov2017bound}
\bibinfo{author}{Bulgakov, E.~N.} \& \bibinfo{author}{Maksimov, D.~N.}
\newblock \bibinfo{title}{Bound states in the continuum and polarization
  singularities in periodic arrays of dielectric rods}.
\newblock \emph{\bibinfo{journal}{Physical Review A}}
  \textbf{\bibinfo{volume}{96}}, \bibinfo{pages}{063833}
\newblock  (\bibinfo{year}{2017}).

\bibitem{yuan_strong_2017}
\bibinfo{author}{Yuan, L.} \& \bibinfo{author}{Lu, Y.~Y.}
\newblock \bibinfo{title}{Strong resonances on periodic arrays of cylinders and
  optical bistability with weak incident waves}.
\newblock \emph{\bibinfo{journal}{Physical Review A}}
  \textbf{\bibinfo{volume}{95}}, \bibinfo{pages}{023834}
\newblock  (\bibinfo{year}{2017}).

\bibitem{liang_three-dimensional_2012}
\bibinfo{author}{Liang, Y.}, \bibinfo{author}{Peng, C.},
  \bibinfo{author}{Sakai, K.}, \bibinfo{author}{Iwahashi, S.} \&
  \bibinfo{author}{Noda, S.}
\newblock \bibinfo{title}{Three-dimensional coupled-wave analysis for
  square-lattice photonic crystal surface emitting lasers with
  transverse-electric polarization: finite-size effects}.
\newblock \emph{\bibinfo{journal}{Optics Express}}
  \textbf{\bibinfo{volume}{20}}, \bibinfo{pages}{15945--15961}
\newblock  (\bibinfo{year}{2012}).

\bibitem{wang_mode_2016}
\bibinfo{author}{Wang, Z.} \emph{et~al.}
\newblock \bibinfo{title}{Mode splitting in high-index-contrast grating with
  mini-scale finite size}.
\newblock \emph{\bibinfo{journal}{Optics Letters}}
  \textbf{\bibinfo{volume}{41}}, \bibinfo{pages}{3872--3875}
\newblock  (\bibinfo{year}{2016}).

\bibitem{ni_analytical_2017}
\bibinfo{author}{Ni, L.}, \bibinfo{author}{Jin, J.}, \bibinfo{author}{Peng, C.}
  \& \bibinfo{author}{Li, Z.}
\newblock \bibinfo{title}{Analytical and statistical investigation on
  structural fluctuations induced radiation in photonic crystal slabs}.
\newblock \emph{\bibinfo{journal}{Optics Express}}
  \textbf{\bibinfo{volume}{25}}, \bibinfo{pages}{5580--5593}
\newblock  (\bibinfo{year}{2017}).

\bibitem{regan_direct_2016}
\bibinfo{author}{Regan, E.~C.} \emph{et~al.}
\newblock \bibinfo{title}{Direct imaging of isofrequency contours in photonic
  structures}.
\newblock \emph{\bibinfo{journal}{Science Advances}}
  \textbf{\bibinfo{volume}{2}}, \bibinfo{pages}{e1601591}
\newblock  (\bibinfo{year}{2016}).

\bibitem{lu_topological_2014}
\bibinfo{author}{Lu, L.}, \bibinfo{author}{Joannopoulos, J.~D.} \&
  \bibinfo{author}{Solja\v{c}i\'{c}, M.}
\newblock \bibinfo{title}{Topological photonics}.
\newblock \emph{\bibinfo{journal}{Nature Photonics}}
  \textbf{\bibinfo{volume}{8}}, \bibinfo{pages}{821--829}
\newblock  (\bibinfo{year}{2014}).

\bibitem{ozawa2018topological}
\bibinfo{author}{Ozawa, T.} \emph{et~al.}
\newblock \bibinfo{title}{Topological photonics}.
\newblock \emph{\bibinfo{journal}{arXiv preprint arXiv:1802.04173}}
\newblock  (\bibinfo{year}{2018}).

\bibitem{wang_observation_2009}
\bibinfo{author}{Wang, Z.}, \bibinfo{author}{Chong, Y.},
  \bibinfo{author}{Joannopoulos, J.~D.} \& \bibinfo{author}{Solja\v{c}i\'{c},
  M.}
\newblock \bibinfo{title}{Observation of unidirectional backscattering-immune
  topological electromagnetic states}.
\newblock \emph{\bibinfo{journal}{Nature}} \textbf{\bibinfo{volume}{461}},
  \bibinfo{pages}{772--775}
\newblock  (\bibinfo{year}{2009}).

\bibitem{luchansky2011high}
\bibinfo{author}{Luchansky, M.~S.} \& \bibinfo{author}{Bailey, R.~C.}
\newblock \bibinfo{title}{High-q optical sensors for chemical and biological
  analysis}.
\newblock \emph{\bibinfo{journal}{Analytical chemistry}}
  \textbf{\bibinfo{volume}{84}}, \bibinfo{pages}{793--821}
\newblock  (\bibinfo{year}{2011}).

\bibitem{di2009chemical}
\bibinfo{author}{Di~Falco, A.}, \bibinfo{author}{O’faolain, L.} \&
  \bibinfo{author}{Krauss, T.}
\newblock \bibinfo{title}{Chemical sensing in slotted photonic crystal
  heterostructure cavities}.
\newblock \emph{\bibinfo{journal}{Applied physics letters}}
  \textbf{\bibinfo{volume}{94}}, \bibinfo{pages}{063503}
\newblock  (\bibinfo{year}{2009}).

\bibitem{logan2018400}
\bibinfo{author}{Logan, A.~D.} \emph{et~al.}
\newblock \bibinfo{title}{400\%/w second harmonic conversion efficiency in
  $14\mu m$-diameter gallium phosphide-on-oxide resonators}.
\newblock \emph{\bibinfo{journal}{arXiv preprint arXiv:1810.06393}}
\newblock  (\bibinfo{year}{2018}).

\bibitem{hirose2014watt}
\bibinfo{author}{Hirose, K.} \emph{et~al.}
\newblock \bibinfo{title}{Watt-class high-power, high-beam-quality
  photonic-crystal lasers}.
\newblock \emph{\bibinfo{journal}{Nature photonics}}
  \textbf{\bibinfo{volume}{8}}, \bibinfo{pages}{406}
\newblock  (\bibinfo{year}{2014}).

\end{thebibliography}
\bibliographystyle{naturemag}


\begin{addendum}
 \item[Competing Interests] The authors declare that they have no
competing financial interests.
 \item[Correspondence] Correspondence and requests for materials
should be addressed to Chao Peng.~(email: pengchao@pku.edu.cn).
\end{addendum}

\clearpage
\begin{figure}
\centering
\includegraphics[width=8.9cm]{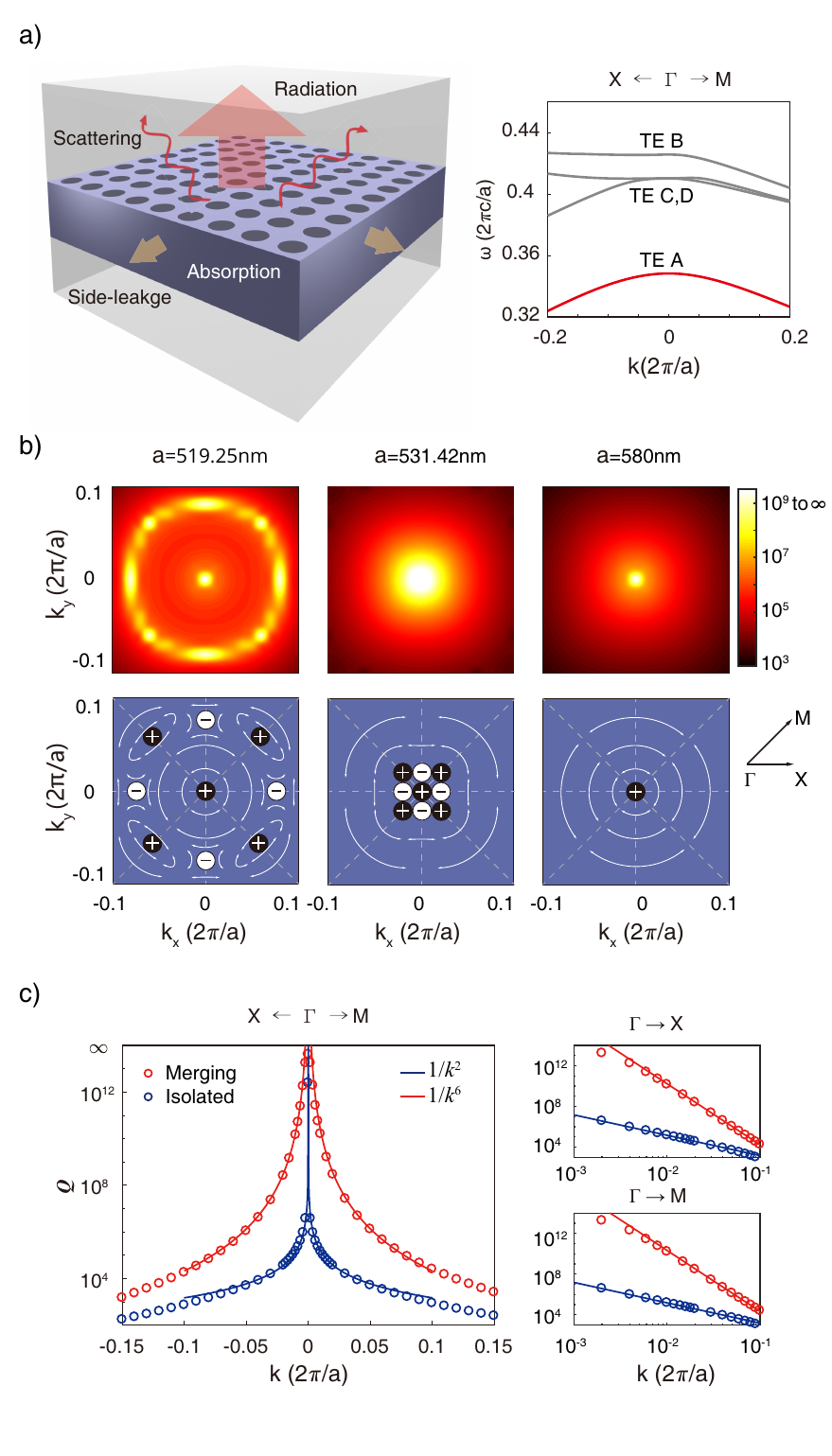}
\caption{{\bf $\mid$ Suppressing radiation losses by merging multiple topological charges of bound states in the continuum (BICs)}.{\bf a,} Schematic of the PhC slab. Band TE-A is marked with a red line.
{\bf b,} Multiple BICs appear on band TE-A, where the normalized radiative lifetime $Q$ diverge.
When sample periodicity $a$ is tuned from 519.25nm to 580nm, 9 BICs with $\pm 1$ topological charges merge into an isolated one with charge $+1$.
{\bf c,} Plots of $Q$ near the center of the BZ when charges just merge (red {$a$=531.42nm}) and long after they have merged (blue {$a$=580nm}). The merging sample (red) shows significantly higher $Q$s than the isolated sample due to a different scaling of $Q \propto 1/k^6$ instead of $Q \propto 1/k^2$, which is observed along both $\Gamma-X$ and $\Gamma-M$ direction. {Simulations here are using FEM.} }
\end{figure}

\clearpage
\begin{figure}
\centering
\includegraphics[width=18.3cm]{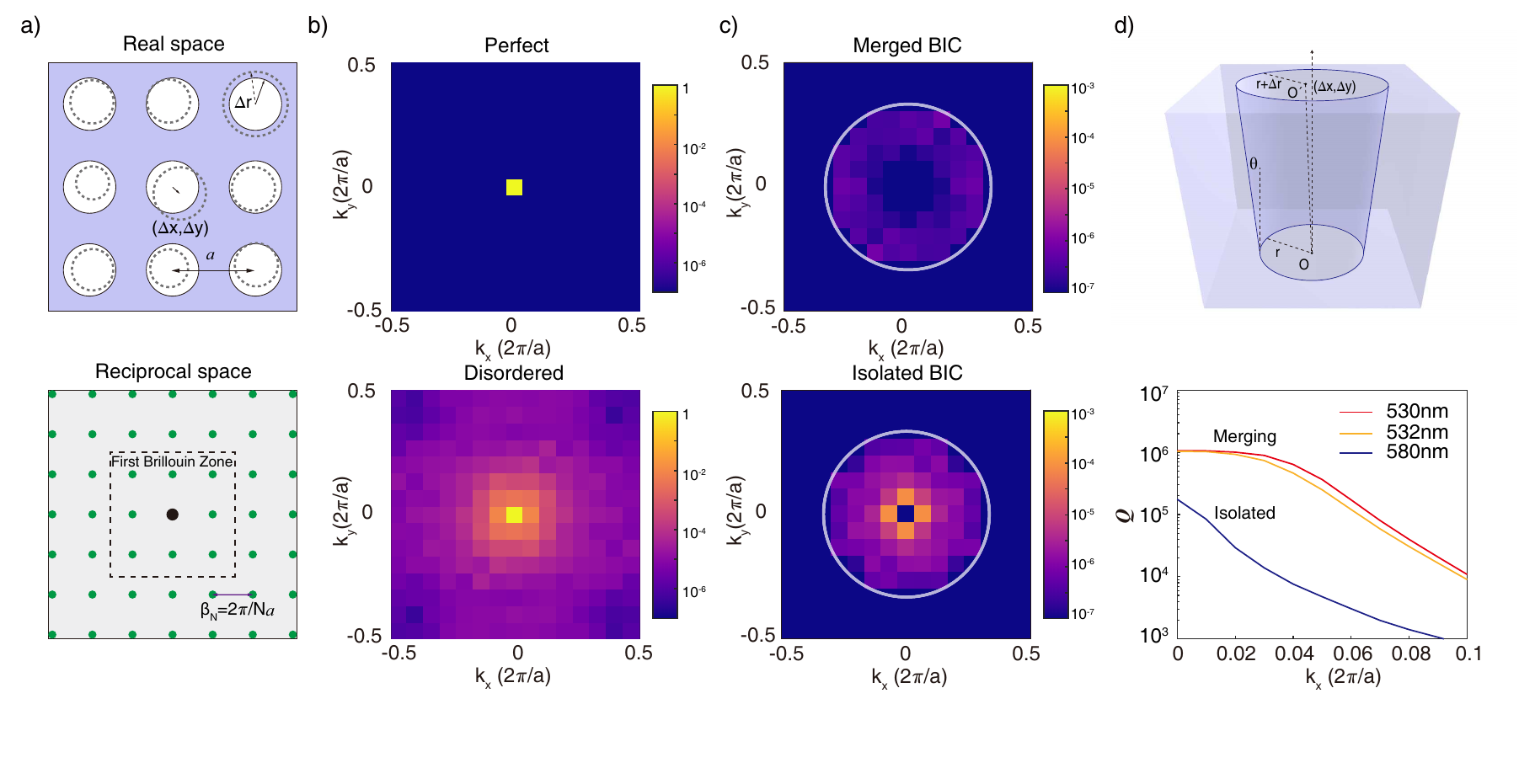}
\caption{{\bf $\mid$ Robustness to scattering losses due to topological protection.}
{\bf a,} Schematic of a fabricated PhC sample (solid lines) with disorder in hole locations and radii compared to a perfect one (dashed lines). Fractional orders of momentum (green dots) are introduced by the super-cell.
{\bf b,} Energy distribution of the highest-Q mode on TE-A band in the momentum space of a merging-BIC design in a perfect (upper) and a disordered structure (lower) inside the first BZ.
{\bf c,} Momentum {energy} distribution of the far-field radiation in a disordered sample with merging BICs (upper) and one with an isolated BIC (lower). The while circles represent the light-cone. The radiative scattering loss is significantly lower in the merging sample than the isolated one. {Simulations are performed in 15$\times$15 super-cell using FEM. {\bf d,} Schematic of typical fabrication error in asymmetric hole (upper) and plots of $Q$ near the BZ center with disorders accordingly (lower).}  }
\end{figure}

\clearpage
\begin{figure}
\centering
\includegraphics[width=8.9cm]{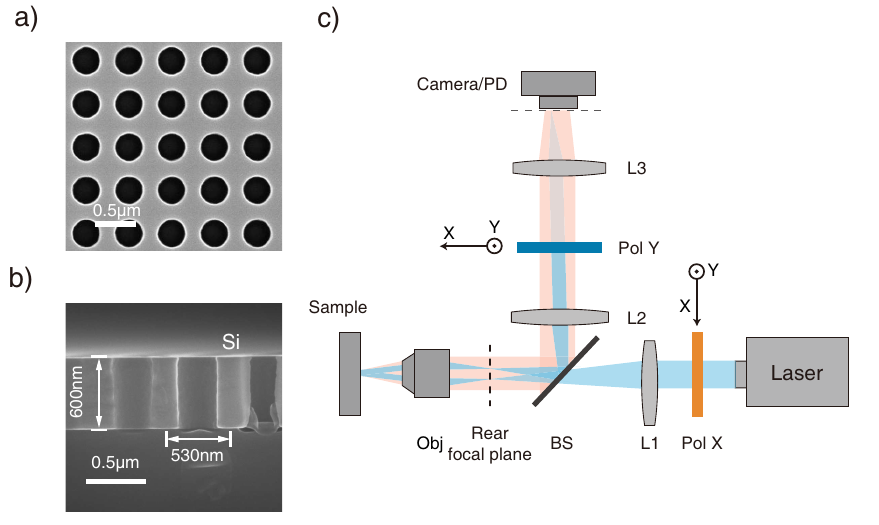}
\caption{{\bf $\mid$ Experimental setup.}
{\bf a,b} Scanning electron microscope (SEM) images of the fabricated PhC sample from the top and cross-section view. The chosen structural parameters correspond to when the 9 BICs just merge in middle panel of Fig. 1b.
The underlying SiO$_2$ layer is later removed for measurements.
{\bf c,} Schematic of the measurement setup. The blue lines refer to the incident light and its direct reflection. The red region refers to radiation losses induced by scattering from disorder. L, lens; Obj, objective; PD, photodiode; Pol, polarizer. }
\end{figure}

\clearpage
\begin{figure}
\centering
\includegraphics[width=18.3cm]{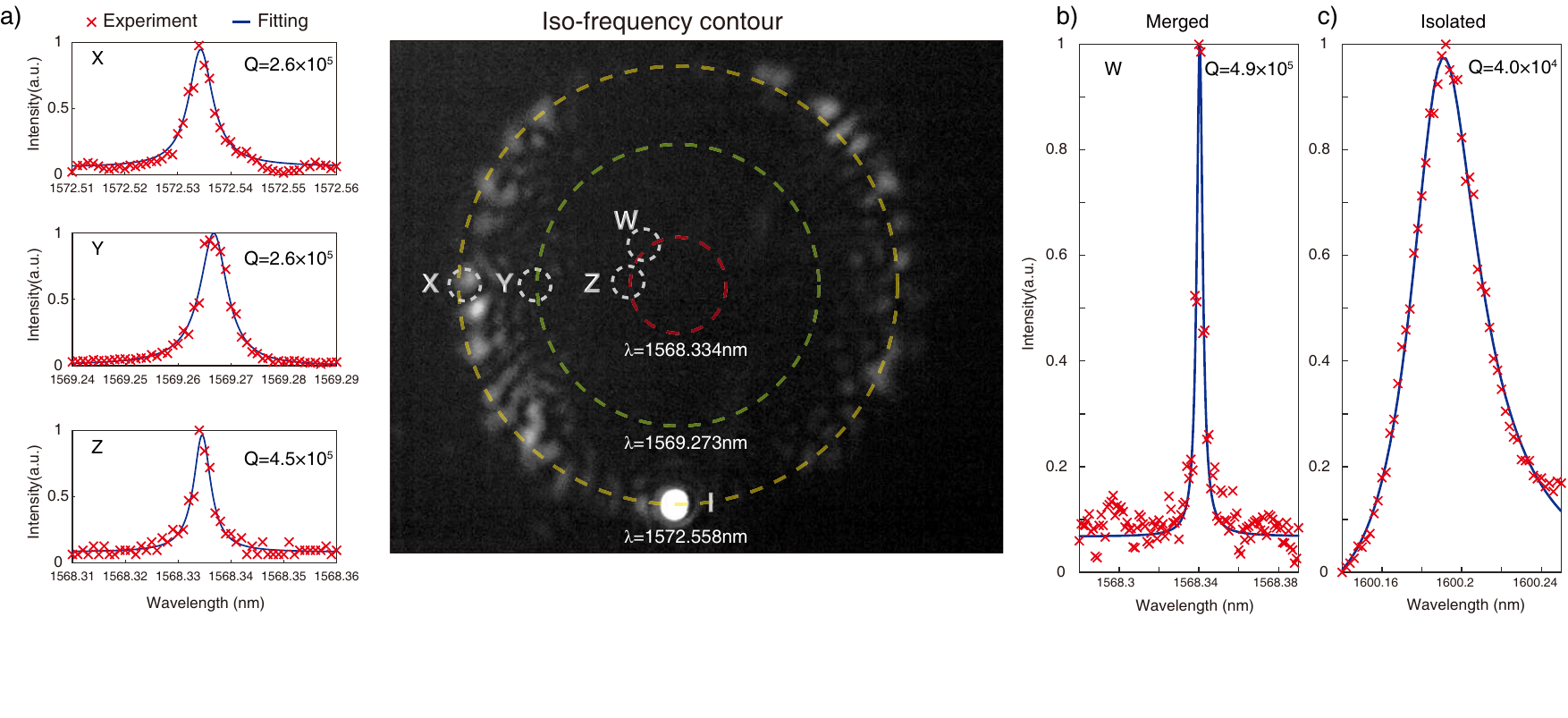}
\caption{{\bf $\mid$ Experimental results.} {\bf a,}
Iso-frequency contours of the sample at different wavelengths are observed on the camera.
Three examples at $1572.558$ nm, $1569.273$ nm, and $1568.334$ nm are shown as dashed lines.
Scattered light intensity at different points in the momentum space (X,Y,Z) are further characterized by a PD, which all exhibit symmetric Lorentzian functions as the incident wavelength. The linewidth is determined by the $Q$ of the underlying resonance.
{\bf b,} The highest $Q$ observed in the merging-BIC sample is $4.9\times 10^5$ at point W, which is over an order of magnitude than the isolated-BIC sample fabricated under the same processes ($Q = 4.0\times 10^4$ as shown in {\bf c}). }
\end{figure}

\clearpage
\begin{figure}
\centering
\includegraphics[width=8.9cm]{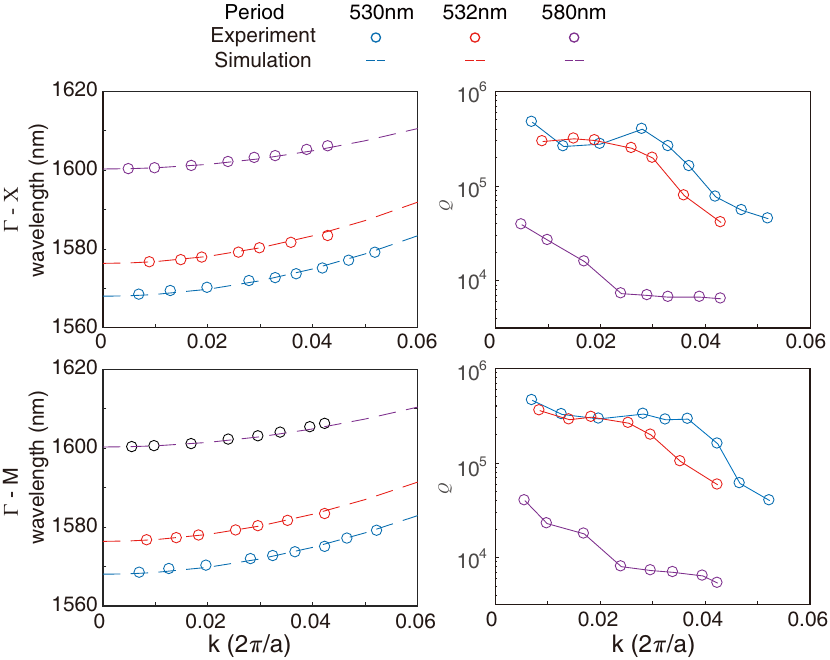}
\caption{{\bf $\mid$ Twelve-times enhancement of quality factors via topological protection.}
{\bf a,} The dispersion of resonances are measured at different points in the momentum space (circles), which show good agreements with simulation predictions with FEM (dashed lines) both along $\Gamma-X$ (upper panel) and $\Gamma-M$ directions (lower panel).
{\bf b,} Over ten-fold enhancement of $Q$ is observed over a wide range in the momentum space in the merging-BIC samples (red and blue) compared to the isolated-BIC sample (purple) due to topological protection.}
\end{figure}
\end{document}